\documentclass[a4paper,11pt]{article}
\pdfoutput=1
\usepackage{jheppub}
\usepackage[T1]{fontenc}
\usepackage{amsfonts}
\usepackage{amsmath}
\usepackage{amssymb}
\usepackage{multirow}
\usepackage{graphicx}
\usepackage{mathtools}
\usepackage{comment}

\usepackage{caption}
\usepackage{subcaption}

\numberwithin{equation}{section}

\begin{document}

\title{Causality bounds from charged shockwaves in 5d}

\author[a]{Sera Cremonini,}
\emailAdd{cremonini@lehigh.edu}
\affiliation[a]{Department of Physics, Lehigh University, Bethlehem, PA, 18015, USA}
\author[b,c]{Brian McPeak,}
\emailAdd{bmmcpeak@syr.edu}
\affiliation[b]{Department of Physics, McGill University}
\affiliation[c]{Department of Physics, Syracuse University}

\author[a]{Mohammad Moezzi,}
\emailAdd{mom323@lehigh.edu}

\author[a]{Muthusamy Rajaguru}
\emailAdd{muthusamy.rajaguru@lehigh.edu}

\date{\today}

\abstract{Effective field theories are constrained by the requirement that their constituents never move superluminally on non-trivial backgrounds. In this paper, we study time delays experienced by 
photons propagating on charged shockwave backgrounds in five dimensions. 
In the absence of gravity -- where the shockwaves are electric fields sourced by boosted charges -- we derive positivity bounds for the four-derivative corrections to electromagnetism, reproducing previous results derived from scattering amplitudes. 
By considering the gravitational shockwaves sourced by Reissner-Nordstr\"om black holes, we derive new constraints in the presence of gravity. We observe the by-now familiar weakening of positivity bounds in the presence of gravity, but without the logarithmic divergences present in 4d. We find that the strongest bounds appear by examining the time delay near the horizon of the smallest possible black hole, and discuss on the validity of the EFT expansion in this region. We comment on our bounds in the context of the swampland program as well as their relation with the positivity bounds obtained from dispersion relations.}

\maketitle

\section{Introduction}

Causality -- the requirement that causes precede their effects -- is among the most fundamental laws of physics. 
 
Our entire mechanistic understanding of the physical world 

hinges on this basic requirement.
Combined with Lorentz invariance, causality gives a powerful prediction: signals cannot propagate faster than light. If they could, then a Lorentz transformation could change the order of events -- a message could be received before it is sent.
In bottom-up approaches to physics, where one explores which theories are consistent with various basic axioms, 
 causality remains a powerful tool because many types of interactions can lead to causality violations \cite{Adams:2006sv}. 
 
 When phrased in terms of effective field theory (EFT), this can be stated as follows: massless particles in interacting theories will have propagation that is slightly superluminal or slightly subluminal depending on the \textit{sign} of certain higher-derivative interactions. For instance, in the effective theory of electromagnetism in four dimensions (4d), 
described by
\begin{align}
\label{QFTlag4d}
    \mathcal{L} = -\frac{1}{4} F^2 + a_1 (F^2)^2 + a_2 (F \tilde F)^2 \, ,
\end{align}
 causality requires \cite{Adams:2006sv} that the Wilson coefficients $a_1,a_2$ obey
\begin{align}
    \label{4dEM}
     a_1>0 \, , \qquad  a_2>0 \,.
\end{align}
Equation~\eqref{4dEM} was derived by considering photon propagation on translationally invariant backgrounds, which are a small subset of the full class of backgrounds that one could consider. Indeed, an important step forward was made in \cite{Camanho:2014apa} by considering shockwave backgrounds instead of constant ones. Shockwaves are gravitational backgrounds that result from ultra-relativistic boosts of ordinary black holes. They have the simplifying feature that the length contraction from the boost flattens the gravitational field of the massive body. An object traveling past the shockwave feels no interaction except at a single point, and the effect of passing by the shockwave -- the time delay and deflection angle -- are calculable in the infinite-boost limit. Causality dictates that the time delay must be \textit{positive}, 
ensuring that it represents a delay rather than a time advance.
This requirement gives a constraint on higher-derivative interactions in exactly the same way as superluminal propagation on constant backgrounds. 

Recently some of us explored generalizations of this idea in  \cite{Cremonini:2023epg}. 
Just like in theories of gravity, shockwaves in field theory can be obtained by boosting the solutions corresponding to various point sources. 
This allows for a new derivation of the bounds \eqref{4dEM} using positive time delays on shockwave backgrounds. 
Moreover, by boosting Reissner-Nordstr\"om black holes instead of Schwarzschild, we were able to derive bounds on all of the operators in Einstein-Maxwell theory, as we shall review in section~\ref{sec:4dreview}. One outcome of our analysis was that the presence of gravity \emph{weakens} the positivity bounds which apply in pure field theory, because it causes a universal time delay that can compensate for -- and allow -- a slight time advance from the other interactions. This was already understood by \cite{Camanho:2014apa}, and its significance in the context of dispersion relation bounds has been discussed recently in \cite{Caron-Huot:2021rmr, Henriksson:2022oeu, Caron-Huot:2022ugt,Caron-Huot:2022jli, McPeak:2023wmq}.

The purpose of this paper is to generalize the work \cite{Cremonini:2023epg} to five dimensions (5d). There are several reasons for wanting to do this. First of all, \cite{Cremonini:2023epg}, and indeed all related work on causality bounds in 4d gravity, suffer from the problem of IR divergences. These appear in the time delay as log terms, i.e. $\log \rho$, where $\rho$ is the distance from the source. Making the log dimensionless requires that we introduce an IR cutoff, and the time delay diverges as this cutoff is taken to infinity. Thus, while time delays can be derived in 4d, IR divergences complicate their physical interpretation. 

Another motivation for studying bounds on Einstein-Maxwell theory comes from the black hole Weak Gravity Conjecture (WGC), first discussed in \cite{Kats:2006xp}  and later in \cite{Cheung:2014ega, Hamada:2018dde, Cremonini:2019wdk, Loges:2019jzs, Aalsma:2020duv, Cremonini:2020smy} -- see also \cite{Harlow:2022gzl} for a review. 
The idea behind it is that, if a certain combination of EFT coefficients is positive, the higher-derivative terms in the Lagrangian will correct the low-energy gravitational action in such a way that black holes will be slightly superextremal, automatically satisfying the WGC \cite{Arkani-Hamed:2006emk}. Our present results are not enough to establish the black hole WGC, due to the existence of $R_{abcd} R^{abcd}$ terms in 5d. These terms are not probed by the photon equations of motion. We shall comment more on this issue in the discussion.

The results of this paper are part of a growing body of work that has gone into extracting constraints on physical theories from causality \cite{Hartman:2015lfa, Camanho:2016opx, Goon:2016une, Hinterbichler:2017qyt, deRham:2020zyh, AccettulliHuber:2020oou, Bellazzini:2021shn, deRham:2021bll, Chen:2021bvg, Serra:2022pzl, CarrilloGonzalez:2023rmc, Chen:2023rar}. One recent approach, especially in \cite{CarrilloGonzalez:2022fwg, CarrilloGonzalez:2023cbf}, was to systematically consider an array of different backgrounds, in order to determine the strongest possible bounds. Those papers considered spherically symmetric, time-independent backgrounds and can be thought of as complementary to this paper.

A closely related paradigm is to bound EFT coefficients using dispersion relations. In that case, bounds on EFT coefficients are derived using basic properties obeyed by amplitudes, including unitarity, analyticity, and Regge boundedness. This alternative point of view was emphasized alongside causality in \cite{Adams:2006sv} and was in fact well-known before then \cite{Pham:1985cr, Pennington:1994kc,  Ananthanarayan:1994hf, Comellas:1995hq, Dita:1998mh}. Since then a vast amount of work has gone into determining the positivity bounds in a wide range of theories and scenarios (see \cite{Manohar:2008tc, Mateu:2008gv, Nicolis:2009qm, Baumann:2015nta, Bellazzini:2015cra, Bellazzini:2016xrt, Cheung:2016yqr, Bonifacio:2016wcb,  Cheung:2016wjt, deRham:2017avq, Bellazzini:2017fep, deRham:2017zjm, deRham:2017imi, Bonifacio:2017nnt, Bellazzini:2017bkb, Bonifacio:2018vzv, deRham:2018qqo, Zhang:2018shp, Bellazzini:2018paj, Bellazzini:2019xts, Melville:2019wyy, deRham:2019ctd, Alberte:2019xfh, Alberte:2019zhd, Bi:2019phv, Remmen:2019cyz, Ye:2019oxx, Herrero-Valea:2019hde, Zhang:2020jyn, Trott:2020ebl,Zhang:2021eeo, Wang:2020jxr, Li:2021lpe, Du:2021byy, Davighi:2021osh, Chowdhury:2021ynh, Henriksson:2021ymi, Caron-Huot:2021enk, Caron-Huot:2022jli, Caron-Huot:2022ugt, Bern:2021ppb, Henriksson:2022oeu, Fernandez:2022kzi, Albert:2023jtd,Bellazzini:2023nqj, McPeak:2023wmq, Bertucci:2024qzt} for an incomplete set). 

The dispersion relation bounds are believed to enforce causality in some manner, as causality ultimately underlies the crucial assumption that the amplitudes are analytic.
However, it remains unclear whether the bounds derived from dispersion relations are the same as those obtained from causality in non-trivial backgrounds.
Indeed, comparing the two sets of bounds was one of the primary motivations of \cite{CarrilloGonzalez:2022fwg, CarrilloGonzalez:2023cbf}. The results seem to be that the causality bounds known so far are strictly \emph{weaker} than the dispersive bounds. However, while there exists a numerical recipe \cite{Arkani-Hamed:2020blm, Bellazzini:2020cot, Tolley:2020gtv, Caron-Huot:2020cmc, Sinha:2020win} for obtaining the strongest possible dispersive bounds, there is no analogous result for causality bounds. Thus, the question remains open. 

Another motivation of this work is to address a gap in the dispersion relation literature for spinning particles in $d>4$. The dispersion relation methods for scalars can be easily generalized to higher dimensions by replacing Legendre polynomials with Gegenbauer polynomials. However, for spinning particles, the group theory determining the kinematic invariants is significantly more complicated in $d > 4$ and was only understood recently \cite{Caron-Huot:2022jli, Buric:2023ykg}. To date, this has not been applied to the Maxwell theory. The leading coefficients to Maxwell theory in $d > 4$ have been bounded in \cite{Hamada:2018dde} by considering scattering that occurs in a 4d subspace. However this approach cannot capture the full set of constraints implied by the 5d symmetry, and no bounds beyond the forward limit have been obtained. In this paper we also only consider the leading (four-derivative) operators but we hope that the techniques used here will eventually allow for bounds on six- and higher-derivative operators.

\paragraph{Results and limits} The results of this paper are the time delays experienced by photons polarized either parallel or perpendicular to the impact parameter, traveling on backgrounds sourced by boosted charged black holes. 
When computing the time delays we have worked with general values of the impact parameter. However, for simplicity we present most of our results in a number of limits, in which the expressions become tractable: large impact parameter $\rho$, near the horizon (or, more precisely, as close to it as the EFT cutoff will allow) and what we call the ``scaling limit'' where $m$ and $q^2$ approach zero at infinite boost.

Let us comment on the meaning of the scaling limit. In their original paper \cite{Aichelburg} on obtaining shockwave metrics from boosted (Schwarzschild) black holes, Aichelburg and Sexl scaled the mass down with the boost, defining a new ``mass'' $m_0 = \gamma m $ which is constant in the infinite boost $\gamma \to \infty$ limit. This convention has the obvious advantage that it keeps the stress-tensor of the spacetime finite -- an infinitely boosted finite mass black hole will have infinite energy. Later, authors \cite{Lousto:1988ua, Lousto:1989ha, Lousto:1990wn, Cremonini:2023epg} adopted and expanded this convention to deal with charged black holes, by taking $q_0^2 = \gamma q^2 $. Of course, 
the black hole charge, when viewed as parametrizing solutions to \emph{classical} gravity/EFT, can be arbitrarily small, so this choice is fine. However, since we are ultimately interested in quantum gravity, where the charge is expected to be quantized, it is useful to be able to go beyond this limit.

The resulting time delays are infinite. Our view about this is the following: by considering backgrounds outside the scaling limit, where the mass and charge are finite arbitrary quantities, we are allowing the boost (and total energy) to approach infinity in the ultraboost limit. While this leads to an infinite time delay, there is still a causality bound on the sign of that divergence. Said another way, we can always boost enough that only the diverging part of the time delay matters --  that is the case our bounds apply to. We note that this particular diverging piece is not necessarily the same as what is obtained by the scaling limit. For instance, it can contain terms which are non-linear in $m$ and $q^2$, which would die off in the scaling limit. Thus, our analysis is a generalization of what has been considered before. As far as we know, this is the first time that these ``fully general" time delays (not simplified by the scaling limit) have been considered in the literature.

\subsection{Review of shockwaves and time-delays in 4d}
\label{sec:4dreview}

For the reader who wants a bit more context without the full details of the calculation, we will present a schematic review of the results in 4d. 

\paragraph{Shockwaves}

Shockwave solutions were first derived by Aichelburg and Sexl \cite{Aichelburg} by 
boosting a  Schwarzschild black hole to ultrarelativistic speeds. The resulting metric is 
\begin{align}
    ds^2 \ = \ \eta_{\mu \nu} dx^\mu dx^\nu - 4 \, G \, m_0 \log\rho^2 \, \delta(u) \,  du^2 \, ,
\end{align}
where the $du^2$ term describes the shockwave profile, which is traveling on an otherwise flat background. The shockwave is localized at $u = 0$ and is sourced by a particle of mass $m = m_0 / \gamma$ moving in the $x$-direction at the speed of light. In addition, $\rho$ is the transverse distance away from the source, i.e. $\rho = \sqrt{y^2+z^2}$ in four dimensions. The logarithm is particular to 4d and is replaced by $\rho^{4-d}$ in higher dimensions.

\paragraph{Time delays}

One can see that traveling across the shockwave at $u = 0$ will cause a time delay by computing the equations of motion of a probe field propagating on this background. Consider a general background of shockwave form, given by 
\begin{align}
\label{flatsw}
    ds^2= \eta_{\mu \nu} dx^\mu dx^\nu +h(u,x_i)du^2\,,
\end{align}
where the shockwave is described by the profile function $h$. Then the wave equation for a scalar probe  is given by 
\begin{align}
\label{scalareom}    \nabla^2\phi=\partial_u\partial_v\phi+h\partial_v^2\phi-\frac{1}{4}\partial_i^2\phi=0\,\,.
\end{align}
We will consider a probe traveling in the $-x$ direction, so that it does not depend on $y$ and $z$. The equation of motion then becomes
\begin{align}
\partial_v\bigl(\partial_u\phi+h\partial_v\phi\bigr)=0 \, .
\label{eq:scalarEOM2}
\end{align}
Now, if the shockwave profile can be modeled by $h = X \delta(u)$, then the solutions will be normal oscillating solutions to the free equations of motion away from $u = 0$, and discontinuous at $u = 0$. The equations of motion can be solved by 
\begin{align}
    \phi = \phi_\text{free}(u, v - X \theta(u))\, ,
\end{align}
where $\phi_\text{free}$ describes a solution to $\partial_v \partial_u \phi = 0$. 
Thus, we see that any equation of the form~\eqref{eq:scalarEOM2} will describe a freely propagating wave that experiences a time delay $X$ when passing through the shockwave.

\paragraph{Photons on charged shockwaves}

In \cite{Cremonini:2023epg} we went beyond this setup to consider photons propagating on charged shockwave backgrounds. Charged shockwaves result from boosting electric charges -- or Reissner-Nordstr\"om black holes, in the case with gravity -- to the speed of light. The solutions with charge are fairly easy to determine by following the same procedure used by Aichelburg and Sexl. The result is the shockwave solution
\begin{align}
\label{finalRN}
    ds^2 \ = \ \eta_{\mu \nu} dx^\mu dx^\nu - 
    \Biggl(8\,m_0\,\ln{\rho+\frac{3}{2}\pi\,\frac{q_0^2+p_0^2}{\rho}}\Biggr)\delta(u)
    \, du^2 \, .
\end{align}
Here we have allowed the shockwave to have electric charge $q$ and magnetic charge $p$, which are scaled in the infinite boost limit by $q_0^2 = \gamma q^2$, $p_0^2 = \gamma p^2$. Thus, we immediately see from the form of the profile function $h$ that introducing charge changes the time delay, by a factor proportional to $q^2$ and $p^2$. 

Determining the behavior of a photon traveling on this background is a bit more complicated because there are now two polarizations. The exact details in 4d will not be relevant here but we note that it is convenient to choose one polarization to be parallel to $\rho$, 
\begin{align}
    E^1_y = \frac{y}{\rho^2} (\partial_u + \partial_v) \phi_1 \, , \qquad E^1_z = \frac{z}{\rho^2} (\partial_u + \partial_v) \phi_1 \, ,
\end{align}
and the other one orthogonal to it, 
\begin{align}
    E^2_y = \frac{z}{\rho^2} (\partial_u + \partial_v) \phi_2 \, , \qquad E^2_z = -\frac{y}{\rho^2} (\partial_u + \partial_v) \phi_2 \, ,
\end{align}
where $\phi_1$ and $\phi_2$ are undetermined functions of $u$ and $v$ only. What we have essentially done is write an ansatz for the probe photon field strength which simplifies drastically the equations of motion. In terms of these  polarizations, the equations of motion become
\begin{align}
    \partial_u \partial_v \phi + \left[- \frac{3 \pi q_0^2}{2 \rho} - 8 m_0 \log \rho \right] \delta(u) \partial_v^2 \phi \ = \ 0 \, .
\end{align}
We stress that the resulting time delay in this case (\textit{i.e.} without higher-derivatives) is the same for both polarizations, 
and is given by 
\begin{align}
    \Delta v \ = \ - \frac{3 \pi q_0^2}{2 \rho} - 8 m_0 \log \rho \,.
\end{align}
The presence of the $\log \rho$ in this formula is ambiguous without a reference scale to make $\rho$ dimensionless. This is typically handled by introducing an IR cutoff, $r_0$. One might consider a conservative value of the cutoff to be the Hubble radius in our universe (or the Anti de Sitter radius, if this calculation was performed there). However, in exactly flat space there is no principled candidate for the value of the IR cutoff, and formally the time delay is divergent. This is one motivation for considering $d>4$, as we do in this paper. 

\paragraph{Higher-derivative corrections to the time delay} 
The leading higher-derivative corrections to Einstein-Maxwell theory in 4d are 
\begin{align}
\label{introLag}
    \mathcal{L} \ = \ \sqrt{-g} \left( \frac{M_{\text{P}}^2}{2} R -\frac{1}{4} F_{\mu \nu} F^{\mu \nu} + \alpha_1 (F_{\mu \nu} F^{\mu \nu})^2 + \alpha_2 ( F_{\mu \nu} \tilde{F}^{\mu \nu})^2 + \alpha_3 W_{\mu \nu \rho \sigma} F^{\mu \nu} F^{\rho \sigma} \right).
\end{align}
The $R_{\mu \nu \rho \sigma} R^{\mu \nu \rho \sigma}$ can be removed using the fact that the Gauss-Bonnet term is topological and therefore cannot affect the equations of motion. In the presence of these higher-derivative terms, we found that the time delay was
\begin{align}
   \Delta v= -\frac{3 \pi q_0^2}{2 \rho} - 8 m_0 \log\rho   
         +    \alpha_i  \frac{48 \pi q_0^2}{\rho^3} \pm  \,\alpha_3 
 \left(\frac{9}{ \rho^3} - \frac{32 m_0}{ \rho^2} \right).
 \label{eq:intdelaywithmp}
\end{align}
For polarization $1$, $\alpha_i = \alpha_1$, and the $\pm \to +$, while polarization $2$ gives $\alpha_i = \alpha_2$, and $\pm \to -$. This illustrates the utility of suitably chosen polarizations. 
An arbitrary polarization will not have a single time delay: it will have a component (\textit{i.e.} its projection onto polarization 1) whose time delay involves $\alpha_1$, and a component (the one proportional to polarization 2) which involves $\alpha_2$. Our choice of polarizations therefore diagonalizes the $\alpha_1$ and $\alpha_2$ interactions -- each polarization experiences a single time delay. This remains true for $\alpha_3$ in the presence of electric or magnetic charge, but when both charges are present, the two polarizations again rotate into each other and there is no single time delay. Different polarizations would be needed to find a single time delay, and these might depend on the particular values of the Wilson coefficients $\alpha_i$.

\section{Charged shockwaves and time delays in field theory}
\label{QFTsection}

Before considering gravity, we can examine how the requirement of causality on shockwave backgrounds constrains higher-derivative operators in quantum field theory. By ``shockwaves'' we mean highly boosted charged sources -- we will study photon propagation on a background with a boosted Coulomb field. This will be described by five-dimensional Maxwell theory plus four derivative corrections,
\begin{align}
\label{QFTlag5d}
    \mathcal{L}_{5d} = -\frac{1}{4} F^2 + \alpha_1 (F^2)^2 + \alpha_2 F^4 \, ,
\end{align}
where $F^4 = F_{\alpha\beta }F^{\beta \gamma} F_{\gamma \delta} F^{\delta \alpha}$.
We compute the time 
delay experienced by a photon $a_\mu$ which scatters off a charged shockwave in flat space. 
Requiring causality in this theory will then lead to constraints on 
$\alpha_1$ and $\alpha_2$. 

To generate the charged shockwave we take the infinite boost limit of a charged particle in flat space. We start with the vector field profile for a charged particle,
\begin{align}
\label{5dggfield}
    \bar{A} = \sqrt{3} \frac{  q}{r^2} \, dt =  \frac{\sqrt{3}\, q}{x^2+y^2+z^2+w^2} \, dt\, , 
\end{align}
where $\{x,y,z,w\}$ denote the four spatial dimensions, and the bar indicates that this will act as our background gauge field. 
We then boost it along the $x$ direction,
\begin{align}
t\rightarrow \gamma (t-\beta x), \quad x\rightarrow \gamma (x-\beta t) \, ,
\end{align}
yielding
\begin{align}
\label{boostedgaugeQFT}
\bar{A} =  \frac{\sqrt{3} q}{\gamma^2(x-\beta t)^2+y^2+z^2+w^2} \, \gamma (dt-\beta dx) \, .
\end{align}
The shockwave that arises in the infinite boost limit $1-\beta  = \epsilon^2 <<0$ will provide the 
background against which the photon will scatter, i.e. $$A_\mu = \bar{A}_\mu + a_\mu \, .$$ 
Since the boost singles out the $t-x$ plane, these two coordinates will play a special role in the analysis that follows, with the remaining coordinates contributing only in the form of an impact parameter $\rho$, i.e. 
$\rho^2 = y^2+z^2+w^2$.

We start by solving for the photon profile before it interacts with the shockwave and
in the absence of higher-derivative terms. 
In this case, one simply needs to solve Maxwell's equations 
$\partial_\mu f^{\mu\nu} =0$,
where $f_{\mu\nu}=\partial_\mu a_\nu - \partial_\nu a_\mu$.
It is particularly convenient to take the ansatz for the probe gauge field to be of the form 
\begin{equation}
\label{probegaugefield}
a_{\mu} = \Phi(t,x)\Bigl(0,0,a_{y}(y,z,w), a_{z}(y,z,w) ,a_{w}(y,z,w)    \Bigr),
\end{equation}
and work in Lorenz gauge, 
so that $\partial_\mu a^\mu=0$.
Then, for this ansatz, solving $\partial_\mu f^{\mu\nu} =0$ is 
equivalent to requiring that 
 the field $\Phi(t,x)$  obeys the wave equation.

 \begin{equation}
 (\partial_t^2-\partial_x^2)\Phi(t,x)=0 \, ,
 \end{equation} and each $a_i(y,z,w)$ component satisfies Laplace's equation in $d=D-2$ transverse directions $x_i$, 
\begin{equation}
\label{laplace}
  \partial^j \partial_j \; a_i =0   \, , 
\end{equation}
where here $d=3$ and $i=y,z,w$.
Solving (\ref{laplace})
then yields the following 
dependence on the
transverse coordinates,
\begin{equation} \label{ansatzGen}
a_{\mu} = \frac{\Phi(t,x)}{\left( y^{2} + z^{2} + w^{2}\right)^{3/2}}\left(0,0,c_{1} y + c_{2} z + c_{3} w, d_{1} y + d_{2} z + d_{3}w, e_{1}y + e_{2}z + e_{3}w \right),
\end{equation}
with the following relations between the coefficients:
\begin{equation} \label{coeffs}
 e_{3} = d_{2} =c_{1} \, , \quad d_{1} = -c_{2} \, , \quad e_{1} = - c_{3}\, ,  
\quad e_{2} = -d_{3} \, .
\end{equation}
By construction, this solution satisfies the Lorenz gauge.

It is important to note that once the higher-derivative terms are restored, (\ref{ansatzGen}) will no longer be a solution of the full equations of motion. The ansatz for $a_\mu$ would need to be corrected, and in particular, the $a_x$ and $a_t$ components, which here we have set to zero, will no longer vanish (they will receive order $\alpha_i$ corrections).
However, order $\alpha_i$ corrections to the probe will give order $\alpha_i^2$ corrections to the time delay, which we are neglecting.
Thus, for our purposes, we do \emph{not} need the corrected solution,
and (\ref{ansatzGen}) will suffice. Indeed, in what follows we will ignore the back-reaction to (\ref{ansatzGen}).

Next, we include the four-derivative operators. The equation of motion for the probe photon is then given by
\begin{equation} \label{EOMflat}
\partial_{\mu} f^{\mu \nu} =  8 \alpha_{1} \partial_{\mu} (2 \bar F^{\mu \nu} \bar F \cdot f + f^{\mu \nu} \bar F^{2} ) + 8 \alpha_{2} \partial_{\mu} (\bar F^{\nu \rho}f_{\rho \sigma}\bar F^{\sigma \mu} + \bar F^{\nu \rho} \bar F_{\rho \sigma}f^{\sigma \mu} + f^{\nu \rho} \bar F_{\rho \sigma} \bar F^{\sigma \mu}) \, ,
\end{equation}
where $\bar{F}_{\mu\nu} = \partial_\mu \bar{A}_\nu - \partial_\nu \bar{A}_\mu $ refers to the background flux. The time delay $\Delta v$ experienced by the probe as it interacts with the shockwave can be extracted by inspecting the components of (\ref{EOMflat}) corresponding to $\nu = y,z,w$, which are the spatial directions transverse to the propagation direction.
As summarized in the introduction,  
the time delay can be read off from the probe equation of motion when it is in the schematic form 
\begin{equation}
\label{schematicEOM}
    \partial_u \partial_v \phi = - 
    %\alpha_i \; 
    \Delta v \; \delta(u) \, \partial_v^2 \phi \, ,
\end{equation}
where $\Delta v$ will contain the contributions from the higher-derivative terms. 

For the charged shockwave background (\ref{boostedgaugeQFT}) and the ansatz (\ref{ansatzGen}), 
the last three components of (\ref{EOMflat})  
%%
%%%%
can be written in the form
\begin{eqnarray}
\label{flatspacegenEOM}
    \partial_u\partial_v \phi &=& 
-\left[\alpha_2 +   (4 \, \alpha_1   + \alpha_2) \frac{c_1 y}{(c_1 y +c_2 z +c_3 w)}\right]
\frac{30 \pi q_0^2 }{\rho^5} \delta(u) \, \partial_v^2 \phi \, , \nonumber \\
    \partial_u\partial_v \phi &=& 
- \left[\alpha_2 +   (4 \, \alpha_1   + \alpha_2) \frac{ c_1 z}{(-c_2 y +c_1 z +d_3 w)}\right]
\frac{30 \pi q_0^2 }{\rho^5} \delta(u) \, \partial_v^2 \phi \, , \nonumber \\
\partial_u\partial_v \phi &=& 
-\left[\alpha_2 +   (4 \, \alpha_1   + \alpha_2) \frac{ c_1 w}{(-c_3 y -d_3 z + c_1 w)}\right]
\frac{30 \pi q_0^2 }{\rho^5} \delta(u) \, \partial_v^2 \phi \, .
\end{eqnarray}
Notice that the $(4 \alpha_1   + \alpha_2)$ term on the right hand side contains explicit dependence on the transverse coordinates, which prevents us from putting the equations into the simple form of (\ref{schematicEOM}) and extracting a sensible time delay. 
There are two ways around this issue.
The first way is to set $c_1=0$, which eliminates the troublesome term, including any dependence on $\alpha_1$. 
The choice $c_1=0$ corresponds to a photon polarization described by
\begin{equation} \label{flatpol1}
a_{\mu}^{(1)} = \frac{\Phi(t,x)}{\left( y^{2} + z^{2} + w^{2}\right)^{3/2}}\left(0,0,c_{2} z + c_{3} w, -c_{2} y  + d_{3}w, -c_{3}y - d_{3}z  \right),
\end{equation}
which we refer to as ``polarization one,'' singling out the effects of the $\alpha_2 F^4$ term. 
For this choice, the three equations of motion become identical,  
\begin{align}
\boxed{    \partial_u\partial_v \phi +
  \alpha_2  \frac{ 30 \pi q_0^2 }{ \rho^5} \delta(u) \, \partial_v^2 \phi =0\,. }
\end{align}
%%%%%
\if 0
and equations given by (check signs)
\begin{eqnarray}
  & (c_2 z +c_3 w) \left[  \partial_u\partial_v \phi +
  \alpha_2  \frac{ 30 \pi q_0^2 }{\rho^5} \delta(u) \, \partial_v^2 \phi \right] = 0 \, ,\\
  %%
 & (-c_2 y +d_3 w)  \left[  \partial_u\partial_v \phi +
  \alpha_2  \frac{ 30 \pi q_0^2 }{\rho^5} \delta(u) \, \partial_v^2 \phi \right] = 0 \, ,\\
%
& (c_3 y + d_3 z ) \left[  \partial_u\partial_v \phi +
  \alpha_2  \frac{ 30 \pi q_0^2 }{\rho^5} \delta(u) \, \partial_v^2 \phi \right] = 0 \, .
\end{eqnarray}
\fi
%%%%%
The key point to note is that this is precisely of the form of (\ref{schematicEOM}), which allows us to immediately read off the time delay experienced by this particular polarization,
$$ \Delta v^{(1)} = \alpha_2 \frac{ 30 \pi q_0^2 }{ \rho^5} \, , $$
which vanishes when the impact parameter $\rho\rightarrow \infty$ as expected on physical grounds.
We  can immediately conclude that we can constrain the sign of $\alpha_2$, by using causality, and in particular that we need 
\begin{equation}
    \label{QFTconstraint1}
    \alpha_2 \geq 0
\end{equation}
to ensure that $\Delta v$ is indeed a delay and not a time advance. This result agrees with the analogous analysis done in four dimensions.

When $c_1$ is nonzero, the only way to extract
from (\ref{flatspacegenEOM}) 
an  equation of motion of the form (\ref{schematicEOM}) 
is by setting 
$c_2=c_3=d_3=0$, corresponding to the following polarization,
\begin{equation} \label{flatpol2}
a_{\mu}^{(2)} = \frac{\Phi(t,x)}{\left( y^{2} + z^{2} + w^{2}\right)^{3/2}} \; c_1 \left(0,0, y ,   z , w \right).
\end{equation}
This ensures once again that the unwanted 
dependence on the transverse coordinates cancels,  and that the three equations 
take the same form 
\begin{align}
    \boxed{ \partial_u\partial_v \phi +
  (2\alpha_1 +\alpha_2)  \frac{ 60 \pi q_0^2 }{\rho^5} \delta(u) \, \partial_v^2 \phi =0 }
\end{align}
%
%%%%
\if 0
\begin{eqnarray}
 & c_1 y  \left[  \partial_u\partial_v \phi +
  (2\alpha_1 +\alpha_2)  \frac{ 60 \pi q_0^2 }{\rho^5} \delta(u) \, \partial_v^2 \phi \right] = 0   \, ,\\
  %%
  & c_1 z  \left[  \partial_u\partial_v \phi +
  (2\alpha_1 +\alpha_2)  \frac{ 60 \pi q_0^2 }{\rho^5} \delta(u) \, \partial_v^2 \phi \right] = 0  \, , \\
%
 & c_1 w  \left[  \partial_u\partial_v \phi +
  (2\alpha_1 +\alpha_2)  \frac{ 60 \pi q_0^2 }{\rho^5} \delta(u) \, \partial_v^2 \phi \right] = 0  \, .
\end{eqnarray}
\fi
%%%%%
from which we can again extract the time delay 
$$ \Delta v^{(2)} = (2\alpha_1+\alpha_2) \frac{ 60 \, \pi \, q_0^2 }{\rho^5} \, . $$
Positivity of the latter tells us that 
\begin{equation}
\label{QFTconstraint2}
    2 \alpha_1 + \alpha_2 \geq 0 \ .
\end{equation}

The two polarizations we have worked with are the only choices that allow us to bring the full probe equations into the form of  (\ref{ansatzGen}), 
giving a clear way to extract 
the time delays.
Moreover, note that the two polarizations can be identified as being \emph{transverse} and \emph{parallel} to the impact parameter vector 
 $\Vec{\rho} \equiv (0,0,y,z,w)$. Indeed, we can rewrite them as follows,
\begin{equation}
    \begin{split}
     & a_{\mu}^{(1)} = \frac{\Phi(t,x)}{\mid \vec{\rho} \mid^{3}} \; \epsilon^{(1)}, \quad \epsilon^{(1)} \equiv \left(0,0,c_{2} z + c_{3} w, -c_{2} y  + d_{3}w, -c_{3}y - d_{3}z  \right)\, , \\
     &a_{\mu}^{(2)} = \frac{\Phi(t,x)}{\mid \vec{\rho} \mid^{3}} \; \epsilon^{(2)}, \quad \epsilon^{(2)} \equiv  c_1 \left(0,0, y ,   z , w \right) = c_1 \vec{\rho} \, .
    \end{split}
\end{equation}
From these expressions, it is clear that $\epsilon^{(2)}$ is parallel to $ \vec{\rho}$. 
Moreover, one can easily check that $ \epsilon^{(1)} \cdot \epsilon^{(2)} = \; 0$ for any choice of $\lbrace c_1, c_2, c_3,d_3 \rbrace$, and therefore we have
\begin{equation}
    \epsilon^{(1)} \perp \epsilon^{(2)}, \quad \epsilon^{(1)} \perp \vec{\rho}, \quad \epsilon^{(2)} \parallel  \vec{\rho} \, .
\end{equation}
Thus, our two polarizations can be identified with fluctuations that are transverse and parallel to the direction defined by the impact parameter, 
\begin{equation}
    a_\mu^{(1)} \rightarrow a_{\mu}^{\perp}
\quad \text{and} \quad a_\mu^{(2)} \rightarrow a_{\mu}^{\parallel} \, ,
\end{equation}
in the sense defined above.

Finally, we should mention that 
the constraints (\ref{QFTconstraint1}) and (\ref{QFTconstraint2}) on $\alpha_1, \alpha_2$ were also found in \cite{Hamada:2018dde}, and can be shown to be equivalent to the known bounds on the $F^4$ terms arising in the $D=4$ theory (\ref{QFTlag4d}).
Indeed,
as noted by \cite{Hamada:2018dde}, if we focus on scattering processes that occur on a four-dimensional sub-spacetime, we can use the $D=4$ relation
 \begin{equation}
     F^4 = \frac{1}{2}(F^2)^2
     + \frac{1}{4} (F \tilde F )^2
 \end{equation}
 to rewrite our $D=5$ terms as follows:
 \begin{equation}
     \alpha_1 
     (F^2)^2 + \alpha_2 F^4 
     \rightarrow 
     (\alpha_1 + \frac{1}{2}\alpha_2) (F^2)^2 
     + \frac{\alpha_2}{4} (F \tilde F)^2 
     =  a_1 (F^2)^2 
     + a_2 (F \tilde F)^2 \, .
 \end{equation}
 Thus, the $D=5$ constraints we have just found,
\begin{equation}
   \alpha_2\geq 0 \qquad \text{and} \qquad 2 \alpha_1 + \alpha_2 \geq 0 
\end{equation}
reduce to the known $D=4$ constraints  
\begin{equation}
   a_1 \geq 0 \qquad \text{and} \qquad a_2 \geq 0 \, ,
\end{equation}
 when one specialized to such a sub-space.
 However, we should stress that the shockwave causality bounds can be obtained without ever reducing to four dimensions.

\section{Gravitational Shockwaves}
\label{sec:Gravity}

We are now ready to discuss causality in the presence of gravity. 
We will work with the five-dimensional action described by 
\begin{equation} \label{11}
    S= 
    \int \, d^5 x \, \sqrt{-g}    \left[ \frac{M_{\text{P}}^3}{2}  R-\frac{1}{4}F^2 + \alpha_1 (F^2)^2 + \alpha_2 F^4 + \alpha_3 R_{\mu\nu\rho\sigma} F^{\mu\nu} F^{\rho\sigma} \right ] ,
\end{equation}
where $F=dA$ and $F^4 = F_{\alpha\beta }F^{\beta \gamma} F_{\gamma \delta} F^{\delta \alpha}$.
This describes Einstein-Maxwell theory with four derivative operators which are assumed to be perturbatively small.
We have neglected the Riemann squared term $R_{\mu\nu\rho\sigma} R^{\mu\nu\rho\sigma}$ (which should also be included in the EFT description at the four-derivative level)  simply because the photon time delay is not sensitive to it.
Note that the
perturbative Wilson coefficients $\alpha_i$
have the following mass dimensions, 
\begin{equation}
    [\alpha_1] = [\alpha_2] = - 5 \, , \quad [\alpha_3] = -2 \, .
\end{equation}
From now on we are going to set $M_{\text{P}}^3=2$. We will restore units only later on when we discuss the implications of our causality bounds.

The gauge field equation of motion obtained 
from the action (\ref{11}) is given by 
\begin{equation}  \label{12}
    \nabla_{\mu} F^{\mu \nu} = 8 \alpha_{1} \nabla_{\mu}(F^{\mu \nu} F^{2}) + 8 \alpha_{2} \nabla_{\mu} (F^{\nu \rho} F_{\rho \sigma}F^{\sigma \mu}) + 4 \alpha_{3} \nabla_{\mu}(R^{\mu \nu \rho \sigma}F_{\rho \sigma})\, .
\end{equation}
We are interested in 
considering a small fluctuation $a_\mu$ of the background gauge field $\bar{A}_\mu $ supporting a
Reissner-Nordstr\"om charged black hole, i.e. $A_\mu = \bar{A}_\mu + a_\mu$. 
Expanding Maxwell's equations (\ref{12}) to linear order in the fluctuations $f_{\mu\nu}$ of the associated field strength, i.e. 
$F_{\mu\nu}=\bar{F}_{\mu\nu}+f_{\mu\nu}$, we find
\begin{equation} \label{14}
\begin{split}
\nabla_{\mu} f^{\mu \nu} =&  \;8 \,\alpha_{1} \nabla_{\mu} (2 \bar F^{\mu \nu} \bar F \cdot f + f^{\mu \nu} \bar F^{2} ) + 8 \alpha_{2} \nabla_{\mu} (\bar F^{\nu \rho}f_{\rho \sigma}\bar F^{\sigma \mu} + \bar F^{\nu \rho} \bar F_{\rho \sigma}f^{\sigma \mu} + f^{\nu \rho} \bar F_{\rho \sigma} \bar F^{\sigma \mu}) \\
& +  4 \alpha_{3} \nabla_{\mu}( R^{\mu \nu \rho \sigma}f_{\rho \sigma}) \, .
\end{split}
\end{equation}
Thus, these are the equations governing the behavior of the probe photon $a_\mu$ 
propagating in the particular background geometry encoded  by $\bar F^{\mu \nu}$ and $\bar g_{\mu \nu} $. The next step is to boost the geometry to generate a shockwave profile which the photon will interact with.

\subsection{The Shockwave Background}

The metric describing the Reissner-Nordstr\"om black hole in $D=5$ flat space is
\begin{equation}
    ds^{2} = - g(r) dt^{2} + g(r)^{-1} dr^{2} + r^{2}d \Omega_{3}^{2}, \qquad g(r) = 1 - \dfrac{2m}{r^{2}} + \dfrac{q^{2}}{r^{4}} \, , 
\end{equation}
where recall that the horizons are located at 
%%%%
$r_h^2 = m \pm \sqrt{m^2-q^2}.$
%%%%
In order to boost the metric, 
it is convenient to rewrite it first in \emph{isotropic} coordinates $\{t,x,y,z,w\}$. To do so, we introduce a new variable 
\begin{equation}
    \bar{r}=\sqrt{x^2+y^2+z^2+w^2} \, ,
\end{equation}
 related to the original radial coordinate $r$ via
\begin{equation}
r = \bar{r} R^{1/2}\, , \qquad \text{with} \qquad 
R(\bar{r}) = \left( 1 + \dfrac{m}{\bar{r}^{2}} + \dfrac{m^{2} -q^{2}}{4 \bar{r}^{4}} \right)  \, .
\end{equation}
The metric then takes the following form,
\begin{equation}
    \begin{split}
&ds^{2} = - g_{00}(\bar{r}) dt^{2} + g_{11}(\bar{r}) \left( dx^{2} + dy^{2} + dz^{2} + dw^{2}\right), \quad
\text{with} \quad  
g_{11}(\bar{r}) = R(\bar{r}) \, .
\end{split}
\end{equation}
In terms of the new $\bar r$ coordinate,
the outer horizon is located at
\begin{equation}
    \bar{r}_{h}^{2} = \frac{1}{2} \sqrt{m^{2} - q^2} \, .
\end{equation}
Finally, the gauge field supporting the charged black hole is  
\begin{equation}
    A_t = \sqrt{3} \dfrac{q}{ r^{2}}dt = \dfrac{\sqrt{3}  q}{\bar{r}^{2}R(\bar{r})},\\
\end{equation}
where in the second step we have converted to isotropic coordinates.

Next, we perform a boost in the $x$ direction, 
\begin{equation}
\label{boost}
    t \rightarrow \gamma (t - \beta x) \, , \qquad 
        x \rightarrow \gamma (x - \beta t) \, , 
\end{equation}
under which the metric becomes
\begin{equation}
    g_{\mu \nu} = \begin{pmatrix}
    \gamma^{2} \left( - g_{00} + \beta^{2} g_{11} \right) & \beta \gamma^{2} \left(  g_{00} -  g_{11} \right) & 0 & 0 &0\\
     \beta \gamma^{2} \left(  g_{00} -  g_{11} \right) & \gamma^{2} \left( - \beta^{2} g_{00} +  g_{11} \right) & 0 & 0 &0\\
     0 & 0 & g_{11} & 0 & 0\\
     0 & 0 & 0 & g_{11} & 0 \\
     0 & 0  & 0 & 0 & g_{11}
\end{pmatrix},
\end{equation}
where it is understood that
 $ g_{00}$ and $ g_{11}$ are being evaluated on the boosted coordinates.
 Similarly, under the boost (\ref{boost}) the gauge field takes the form
\begin{equation}
    A = \frac{4 \sqrt{3} q \gamma \bar{r}^2}{4 \bar{r}^{4} + 4 m \bar{r}^{2} + m^{2} - q^{2}} \left( dt^{2} - \beta dx^{2} \right) \, .
\end{equation}

While it would be interesting to examine a general (finite) boost, it is technically much more feasible to consider ultra-relativistic speeds. 
Indeed, we are going to work in the 
ultra-relativistic limit $\beta=1-\epsilon^2$, with $\epsilon <<1$. 
This allows us to turn the black hole background into a spacetime geometry which describes a gravitational shockwave,
\begin{align}
\label{swmetric}
    ds^2= \eta_{\mu \nu} dx^\mu dx^\nu +h(u,x^i) \, du^2\,,
\end{align}
where $x^\mu=\{u,v,x^i\}$,
$u=t-x$, $v=t+x$ and the $x^i$ label the coordinates transverse to the $u-v$ plane.
The profile function $h(u,x^i)$ then captures the geometrical properties of the shockwave.
As in Section 2, we introduce an impact parameter $\rho$, defined 
through 
\begin{equation}
  \rho^2 = y^2+z^2+w^2  \, , 
\end{equation}
which measures the distance from the shockwave along the transverse coordinates.
In the ultra-relativistic limit, the shockwave profile then takes the form
\begin{align}
h(u,x_{i}) &=  \frac{1}{\epsilon}\Biggl[\frac{\sqrt{2}\pi(m^{2} - q^{2} + 8 m \rho^{2})}{16 \rho^{3}} + \frac{\pi m^{2}(\tilde\rho_- - \tilde\rho_+ )}{ q ( m + 2 \rho^{2})}  \nonumber \\
& + \frac{\pi}{4} 
(\tilde\rho_+ +\tilde\rho_-)
\left( 1 + \frac{4 \rho^2(m - \rho^2)}{(m + 2 \rho^2)
\tilde\rho_+\tilde\rho_-} \right) 
+ \frac{4 \pi q \rho^{4} }{ (m + 2 \rho^{2})} \left( 
\frac{1}{\tilde\rho_-^3} - \frac{1}{\tilde\rho_+^3}
\right) \Biggr] \, \delta(u) \, , 
\label{generalh}
\end{align}
%%%
where we have introduced the quantities $\tilde{\rho}_{\,\pm} = \sqrt{2 \rho^2+m \pm q}$ 
to write the expression more compactly.
In the infinite boost limit,  
the outer horizon is located at
\begin{equation}
    \rho_{h}^{2} = \frac{1}{2} \sqrt{m^{2} - q^2} \, .
\end{equation}
Thus, we immediately see that for an extremal black hole, the horizon radius vanishes
\begin{equation}
    m=q \quad  \Rightarrow \quad \rho_h^{\text{ext}} = 0 \, .
\end{equation}

The full expressions for the time delays are very complicated so it will be convenient to examine certain limits. Such limiting cases will correspond to the photon probing different regions of the geometry:
\begin{itemize}
    \item {\bf near horizon:} a natural regime to examine will come from taking the impact parameter to be close to the black hole horizon $\rho_h$. 
    We will discuss two ways to approach the horizon region, one corresponding to taking   $\rho =\rho_{h} (1+ \lambda)$ with $\lambda <<1$, and another one given by 
$\rho^2<< m$. As we will see, the first method will be valid for non-extremal black holes, while the second is more appropriate for the extremal case $m=q$. Indeed, while qualitatively the two limits yield similar answers for extremal black holes, they don't commute (the precise numerical factors are different).
\item 
{\bf scaling limit:}
a particularly simple limit of the time delay comes from 
rescaling 
the mass and charge of the black hole with the boost parameter, 
\begin{equation}
\label{scaling}
    m = \frac{m_0}{\gamma} \sim m_0 \sqrt{2} \epsilon\, , \quad  q^2 = \frac{q_0^2}{\gamma} \sim q_0^2 \sqrt{2} \epsilon \, .
\end{equation}
This ``scaling limit'' is special because it leads to a finite total energy even at infinite boost. When there is no charge, the rescaled mass $m_0$ is simply the momentum. However, issues may arise with charge quantization in general, as discussed in \cite{Cremonini:2023epg}. Nonetheless, we may see what is implied by taking it literally.

\item {\bf Large distance}

A final simple limit is the large-distance limit, where we expand around $\rho = \infty$. In this limit, higher-derivative contributions are suppressed by further inverse powers of $\rho$, so that four-derivative terms are $\rho^{-2}$ times smaller than two-derivative terms, and so on. 
\end{itemize}

\subsection{Computing Time Delays}

We shall use the same polarization ansatz (\ref{ansatzGen})-(\ref{coeffs}) we used in the purely field theoretic case, which was given by
\begin{equation}
\label{generalpol}
a_{\mu} = \frac{\Phi(t,x)}{\left( y^{2} + z^{2} + w^{2}\right)^{3/2}}\left(0,0,c_{1} y + c_{2} z + c_{3} w, - c_{2} y + c_{1} z + d_{3}w, - c_{3}y - d_{3}z + c_{1}w \right).
\end{equation}
Recall that we identified the first polarization (corresponding to $c_1=0$) with 
\begin{equation} \label{pol1}
a_{\mu}^{(1)} = \frac{\Phi(t,x)}{\left( y^{2} + z^{2} + w^{2}\right)^{3/2}}\left(0,0,c_{2} z + c_{3} w, -c_{2} y  + d_{3}w, -c_{3}y - d_{3}z  \right),
\end{equation}
while the second (corresponding to $c_2=c_3=d_3=0$ and parallel to $\vec{\rho}$ ) with
\begin{equation} \label{pol2}
a_{\mu}^{(2)} = \frac{\Phi(t,x)}{\left( y^{2} + z^{2} + w^{2}\right)^{3/2}} \; c_1 \left(0,0, y ,   z , w \right).
\end{equation}
Note that $a^{(1)}_\mu$ 
describes more than one polarization, since we have freedom in how we choose the parameters $c_2,c_3,d_3$ 
(this reflects the fact that there is more than one way to construct a vector that is transverse to
$a^{(2)}_\mu$). 
However, as we will see, $a^{(1)}_\mu$ experiences the same time delay  independently of the choice of $c_2,c_3,d_3$. Thus, we treat it as a single polarization.

Let's start by inspecting the structure of 
Maxwell's equations in the absence of higher-derivative corrections.
It is easy to show that, using the probe ansatz (\ref{generalpol}),  
the $\nu=y,z,w$
components
$\nabla_{\mu} f^{\mu \nu}=0$ reduce to the following equation of motion 
\begin{equation}
    \begin{split}
    & \partial_{u} \partial_{v} \Phi - \frac{1}{\epsilon} \Biggl[\frac{\pi}{4} \Biggl(\frac{(m-q)^2}{q  \tilde{\rho}_{-}}-\frac{(m+q)^2}{q \tilde{\rho}_{+}} - \frac{4 m^2 - 2 q^2}{\sqrt{  m^2-q^2} \rho_{H}^{-}}+\frac{4 m^2 - 2 q^2}{\sqrt{m^2-q^2}\ \rho_{H}^{+}}
    \\ & + \frac{q^2 - 2m (m-\sqrt{m^2-q^2}+4 \rho^2)}{ (\rho_{H}^{-})^{3}}
     +\frac{q^2 - 2m (m+\sqrt{m^2-q^2}+4 \rho^2)}{ (\rho_{H}^{+})^{3}} \Biggr) \Biggr] \delta(u)\, \partial_{v}^{2}\Phi
       = 0 \, , \nonumber
    \end{split}
\end{equation}
where we use again $\tilde{\rho}_{\,\pm} = \sqrt{2 \rho^2+m \pm q}$ and have also introduced $\rho_{H}^{\pm} = \sqrt{2 \rho^2 \pm \sqrt{m^2-q^2}}$.
From this equation we can then immediately read off the time delay \emph{due entirely to the geometry} -- the expression in the square brackets above:
\begin{equation}
    \begin{split} \label{geomgeneralcase}
    &  \Delta v^{\,\text{geom}} = - \frac{\pi}{4}  \frac{1}{\epsilon} \Biggl(\frac{(m-q)^2}{q  \tilde{\rho}_{-}}-\frac{(m+q)^2}{q \tilde{\rho}_{+}} - \frac{4 m^2 - 2 q^2}{\sqrt{m^2-q^2} \rho_{H}^{-}}+\frac{4 m^2 - 2 q^2}{\sqrt{m^2-q^2}\ \rho_{H}^{+}}
    \\ & + \frac{q^2 - 2m (m-\sqrt{m^2-q^2}+4 \rho^2)}{ (\rho_{H}^{-})^{3}}
     +\frac{q^2 - 2m (m+\sqrt{m^2-q^2}+4 \rho^2)}{ (\rho_{H}^{+})^{3}} \Biggr) \, . 
    \end{split}
\end{equation}
Since this expression is not particularly illuminating, 
a few limits are worth considering:
\begin{itemize}
\item 
In the scaling limit (\ref{scaling}) the time delay takes the simple form
\begin{equation}
\label{scalinggeom}
    \Delta v^{\,\text{geom}} =   \frac{ \pi (24 m_{0} \rho^{2} - 5 q_{0}^{2} )}{8  \rho^{3}} \, ,
\end{equation}
which agrees with the expression that would be extracted directly from the metric (\ref{metricscalinglimit}) in the same limit.
Indeed, it's easy to show that under (\ref{scaling}) the general 
profile function (\ref{generalh}) simplifies significantly and becomes
\begin{equation}
\label{metricscalinglimit}
     h(u,x_{i}) 
     %= -\frac{\pi (5 q^{2} - 24 m \rho^{2})}{8 \sqrt{2}  \rho^{3}} \delta(u) \,  
     = \frac{\pi (24 m_0 \rho^{2}- 5 q_0^{2}  )}{8  \rho^{3}} \,\delta(u) \, ,
\end{equation}
confirming that in this limit one has
\begin{equation}
    h(u,x_{i})= \Delta v^{\,\text{geom}} \, \delta(u) \, ,
\end{equation}
as we stressed in \cite{Cremonini:2023epg} and noted in the literature.

In order to 
consider arbitrary (and potentially small) values of $\rho$ for generic mass and charge, we cannot work in this scaling limit.
Moreover, 
from (\ref{scalinggeom}) we see that, to ensure a positive time delay, 
we also need  
$24 m_{0} \rho^{2} - 5 q_{0}^{2}\geq 0 $. 
This sets a lower bound on the impact parameter,
$ \rho^{2} \geq \frac{5 q_{0}^{2}}{24 m_{0}}$.

\item If we turn off the black hole charge, we find
\begin{equation}
    \begin{split}
    \Delta v^{\,\text{geom}} &= - \frac{\pi\; m}{4 \epsilon \left( - m^2 + 4  \rho^4 \right)^{3/2}} \biggl[ \left( 3 m^2 + 2  m \rho^2 - 16  \rho^4  \right) \sqrt{-m + 2 \; \rho^2} \\ &+ \left(4  m^2 - 8 m \rho^2 - 32  \rho^4\right)  \sqrt{m + 2 \; \rho^2}\biggr]. 
    \label{unchargedgeom}
    \end{split}
\end{equation}
In the near 
horizon limit 
$ \rho = \rho_h(1+\lambda)$,
with 
$\rho_h=m/\sqrt{2}$ and  $\lambda <<1 $,
this becomes 
\begin{equation}  \label{unchargedgeomnearh}
    \Delta v^{(1)} = \frac{\pi \sqrt{2} \, \sqrt{m}}{4 \epsilon \,  \lambda^{3/2}},
\end{equation}
which we note is divergent at $\lambda = 0$, and grows with the black hole mass, as expected.
\item For an extremal black hole $m=q$, we find 
    \begin{equation}
        \Delta v^{\text{ geom}} = \frac{\pi m \left( (m + 4 \rho^2) \sqrt{m + \rho^2} + 2\rho^3 \right)}{2 \sqrt{2} \epsilon \; \rho^3 \sqrt{m + \rho^2}}
    \end{equation}
     Note that we can now see explicitly that the near horizon region of the extremal black hole can be probed by taking $\rho^2/m <<1$. 
\end{itemize}

Restoring the higher-derivative corrections entails adding the contribution from the right-hand side of 
(\ref{14}).
Without resorting to particular limits, 
the analysis is quite cumbersome and so are the general expressions for the time delays $\Delta v$. 
Thus, in what follows we will only include explicitly specific cases in which the generic expressions for  $\Delta v$ simplify significantly.

%%%%%%%%%%%%%%%%%%%%%%%%%

\subsubsection{Scaling limit}
We start by examining the scaling limit (\ref{scaling}), which we recall corresponds to taking the impact parameter $\rho$ to be much bigger than the scales set by the mass and charge of the black hole.
Working with our general ansatz (\ref{generalpol}), the left-hand side $\nabla_{\mu} f^{\mu \nu}$ of the equations of motion 
for the probe becomes\footnote{We note that these terms come from expanding the equations of motion in the ultraboost limit, and keeping the leading term, which gives the delta function, and the subleading term in the boost parameter.},  
\begin{equation}
    \begin{split}
    & - (c_{3} w + c_{1}y + c_{2} z) \frac{4 }{\rho^3} \left[ \partial_{u}\partial_{v} \Phi + \frac{ \pi (24 m_{0} \rho^{2} -5 q_{0}^{2})}{8  \rho^{3}} \delta(u) \partial_{v}^{2}\Phi \right] \, ,\\
   & -(d_{3} w - c_{2} y + c_{1} z)  \frac{4 }{\rho^3} \left[   \partial_{u}\partial_{v} \Phi + \frac{ \pi (24 m_{0} \rho^{2} -5 q_{0}^{2})}{8  \rho^{3}} \delta(u)  \partial_{v}^{2}\Phi \right] \, ,\\
   & - (c_{1} w - c_{3} y - d_{3} z) \frac{4 }{\rho^3} \left[   \partial_{u}\partial_{v} \Phi + \frac{ \pi (24 m_{0} \rho^{2} -5 q_{0}^{2})}{8  \rho^{3}} \delta(u)  \partial_{v}^{2}\Phi \right]\, ,
    \end{split}
\end{equation}
where we have only included the components $\nu=y,z,w$ of the equations of motion since they suffice to extract the time delay.
\if 0
From these expressions, one would immediately conclude that the time delay due to the background alone is given by
\begin{equation}
    \Delta v = \frac{ \pi (24 m_{0} \rho^{2} -5 q_{0}^{2})}{8 \rho^{3}}
    = \frac{3\pi }{\rho^3}\left[ m_{0} \rho^2 -\frac{5}{24} q_{0}^{2}   \right]\, ,
\end{equation}
in perfect agreement with what would be extracted directly from the shockwave profile $h$ given in (\ref{metricscalinglimit}), once the mass and charge are appropriately rescaled. 
\fi
%
Next, we compute 
the right-hand side of (\ref{14}),
i.e. the contributions from the higher-derivative corrections 
(the $\nu=y,z,w$ components). Working again with (\ref{generalpol}), we have
\begin{equation}
    \begin{split}
    & \frac{ 120  \pi q_{0}^{2}  }{ \rho^{8}} \Bigl[ c_{1} y  \, (4\alpha_{1} + \alpha_2) \, + (c_{3} w +  c_{1} y + c_{2} z) \alpha_{2} \Bigr] \delta(u) \, \partial_{v}^{2}\Phi \, , \\
 & \frac{ 120  \pi q_{0}^{2}  }{ \rho^{8}} \Bigl[ c_{1} z  \, (4\alpha_{1} + \alpha_2) \, + (d_{3} w -  c_{2} y +  c_{1} z) \alpha_{2} \Bigr] \delta(u) \, \partial_{v}^{2}\Phi \, , \\
 & \frac{ 120  \pi q_{0}^{2}  }{ \rho^{8}} \Bigl[ c_{1} w  \, (4\alpha_{1} + \alpha_2) \, + (c_{1} w -  c_{3} y  - d_{3} z) \alpha_{2} \Bigr] \delta(u) \, \partial_{v}^{2}\Phi \, ,
    \end{split}
\end{equation}
while the contributions from $\alpha_3$ are given by
\begin{equation}
    \begin{split}
 & \frac{6\pi}{\rho^8} \alpha_{3} \Bigl[4 c_{1} y (5 q_{0}^{2} - 4 m_{0} \rho^{2}) + ( c_{2} z  + c_{3} w) (8 m_{0} \rho^{2}-5 q_{0}^{2}) \Bigr] \delta(u)  \partial_{v}^{2}\Phi \, ,\\
 & \frac{6\pi}{\rho^8} \alpha_{3} \Bigl[4 c_{1} z (5 q_{0}^{2} - 4 m_{0} \rho^{2}) + (d_{3} w - c_{2} y) (8 m_{0} \rho^{2}-5 q_{0}^{2}) \Bigr] \delta(u)  \partial_{v}^{2}\Phi \, ,\\
 & \frac{6\pi}{\rho^8} \alpha_{3} \Bigl[4 c_{1} w (5 q_{0}^{2} - 4 m_{0} \rho^{2}) - (c_{3} y + d_{3} z ) (8 m_{0} \rho^{2}-5 q_{0}^{2}) \Bigr] \delta(u)  \partial_{v}^{2}\Phi \, .
    \end{split}
\end{equation}
These expressions were derived for generic polarizations. They simplify nicely when we restrict our attention to particular polarizations.
Indeed, for the first polarization ($c_{1} = 0$) the three equations of motion reduce to  
\begin{equation}
    \begin{split}
&    \partial_{u}\partial_{v} \Phi +   \left[\frac{ \pi (24 m_{0} \rho^{2} -5 q_{0}^{2})}{8 \rho^{3}}  + 
\frac{ 3  \pi }{\rho^{5}} 
\left(10 q_0^2 \, \alpha_2 + (4 m_0 \rho^2 - \frac{5}{2} q_0^2)\alpha_3 \right) 
\right]\delta(u)  \partial_{v}^{2}\Phi  = 0 \, .
    \end{split}
\end{equation}
Similarly, for the second polarization ($c_{2} = c_{3} = d_{3} = 0$) the equation of motion reduces to
\begin{equation}
      \partial_{u}\partial_{v} \Phi + 
     \left[\frac{ \pi (24 m_{0} \rho^{2} -5 q_{0}^{2})}{8 \rho^{3}}   + \frac{6\pi}{\rho^5} \Bigl( 10 q_{0}^{2} \, (2\alpha_{1} + \alpha_2)  +  \alpha_{3}  (5 q_{0}^{2} - 4 m_{0} \rho^{2}) \Bigr) \right] \delta(u) \, \partial_{v}^{2}\Phi  =0 \, .
\end{equation}
From these equations, we can immediately extract the time delays experienced by, respectively, the first and second polarizations, 
\begin{equation} \label{delay1}
    \Delta v^{(1)} = 
    \frac{3\pi }{\rho^3}\left[ m_{0} \rho^2 -\frac{5}{24} q_{0}^{2}   \right]
    + \frac{3\pi}{\rho^5} \left[ 10 \,q_{0}^{2} \, \alpha_{2}   
    +  4\left( m_{0} \rho^{2}- \frac{5}{8} q_{0}^{2} \right)\, \alpha_3 \right],
\end{equation}
and
\begin{equation}  \label{delay2}
    \Delta v^{(2)} = 
    \frac{3\pi }{\rho^3}\left[ m_{0} \rho^2 -\frac{5}{24} q_{0}^{2}   \right] + \frac{6 \pi}{\rho^5}
    \left[10 \, q_{0}^{2} \,(2\alpha_1+\alpha_2)- 4\left(m_0 \rho^2 - \frac{5}{4} q_0^2\right) \, \alpha_3\right].
\end{equation}
We emphasize that these time delays are valid
when the mass and charge are scaled to zero with $\rho$ unconstrained.   
Finally, restoring units and using $M$ and $Q$ to denote the black hole mass and charge,
\footnote{For our setup, the mass dimensions of various quantities are as follows,
\begin{equation}
[M] = 1, \quad [Q] = -1/2, 
\quad [\rho] = -1, \quad [\alpha_{1}] = [\alpha_{2}] = -5, \quad  [\alpha_{3}] = -2\,.
\end{equation} 
Recall that we have worked with $M_{\text{P}}^{3} =2$. 
We use dimensional analysis to restore needed factors of $M_p^3/2$.}
the time delays become 
\begin{equation} \label{delay1units}
    \Delta v^{(1)} = \frac{3 \sqrt{2}\pi}{\epsilon \rho^3\,  M_{\text{P}}^3}  \left( M \rho^2 - \frac{5}{24} Q^2 \right)    + \frac{3 \pi}{\sqrt{2}\epsilon \rho^5} \left[ 10 \,Q^{2} \, \alpha_{2}   
    +  \frac{8}{M_{\text{P}}^3} \left( M \rho^{2}- \frac{5}{8} Q^2 \right)
    %\frac{8 M \rho^{2}- 5Q^2 }{M_{\text{P}}^3}
    \, \alpha_3 \right],
\end{equation}
and
\begin{equation}  \label{delay2units}
    \Delta v^{(2)} = 
    \frac{3\sqrt{2}\pi}{\epsilon \rho^3\,  M_{\text{P}}^3}  \left( M \rho^2 - \frac{5}{24} Q^2 \right)   + \frac{3 \sqrt{2} \pi}{\epsilon \rho^5}
    \left[10 \, Q^{2} \,(2\alpha_1+\alpha_2)- 
    \frac{8}{M_{\text{P}}^3} \left(M \rho^{2} - \frac{5}{4} Q^2\right)\, \alpha_3\right].
\end{equation}
Here we see clearly that the higher-derivative terms lead to contributions to the time-delay that are subleading in the impact parameter $\rho$. Note also that the contributions from the geometry and from the $\alpha_3$ operator vanish in the limit in which 
gravity decouples, \textit{i.e.} $M_{\text{P}}\rightarrow \infty$, as they should, and we recover the pure field theory results $\alpha_2\geq 0$, $2\alpha_1+\alpha_2 \geq 0$.
When the black hole is neutral, i.e. $Q=0$, 
the time delays reduce to the simple expressions
    \begin{equation}  \label{neutraldelays}
    \Delta v^{(1)} = \frac{3 \sqrt{2}\pi M }{\epsilon \rho\,  M_{\text{P}}^3}
    \left(1 + 4 \frac{\alpha_3}{\rho^2}\right) \, , \qquad
    \Delta v^{(2)} = 
    \frac{3 \sqrt{2}\pi M }{\epsilon \rho\,  M_{\text{P}}^3} \left(1 -8 \frac{\alpha_3}{\rho^2}\right)   \, ,
\end{equation}
which reproduce the result of \cite{Camanho:2014apa}, who showed that for each polarization the $R_{\mu\nu\rho\sigma} F^{\mu\nu} F^{\rho\sigma}$ contribution to the time delay has a different sign\footnote{For a direct comparison with 
\cite{Camanho:2014apa}, note that our $\alpha_3$ is defined with the opposite sign of $\hat \alpha_2$ in \cite{Camanho:2014apa}.}.

The main lesson of these bounds -- obtained in the scaling limit -- is that we cannot let $\rho$ become too small. If we tried to take it to be small, then the contribution of the higher-derivative terms would become comparable to the leading term, and we would be forced to consider further higher-derivative corrections. Indeed, it is clear from equation~\eqref{neutraldelays} that this happens at order inverse cutoff scale, $\rho \sim 1 / \Lambda$. At smaller distances, the EFT breaks down and knowledge of the UV completion is required.
Furthermore, the scaling limit does not let us consider extremal black holes, since the mass and charge are scaled differently with the boost parameter.

\subsubsection{Large distance limit}

Previous literature has primarily considered the scaling limit, but it is not strictly necessary to have a finite time delay in the $\gamma \to \infty$ limit -- rather it is the leading divergence in that limit that must be positive. Given this, we will look to other regimes where the results simplify enough to be tractable. Two main candidates are large $\rho$ (which is similar to the scaling limit but places no restriction on the relative size of $m$ and $q$) and $\rho$ near the black hole horizon.

First let us consider the large-$\rho$ limit, in the absence of any scaling of the charges. In both cases, the two-derivative term, up to order $\rho^{-5}$, is the same:
\begin{align}
    \Delta v^{\,\text{geom}} = \frac{\sqrt{2} \pi}{ 16 \epsilon \rho^5} (24 m \rho^4 + 5(m^2-q^2) \rho^2 + 6m^3 - 3 m q^2) \, .
\end{align}
With this, the total time delay is 
\begin{align}
    \Delta v^{(1)} = \Delta v^{\,\text{geom}} + \frac{\sqrt{2} \pi}{ \epsilon \rho^5} \left( 15 q^2 \alpha_2 + \left(6 m \rho^2 - \frac{15}{4} q^2 - 6 m^2 \right) \alpha_3 \right) \, ,
\end{align}
and 
\begin{align}
    \Delta v^{(2)} = \Delta v^{\,\text{geom}} + \frac{\sqrt{2} \pi}{ \epsilon \rho^5}  \left(  30 q^2 (2 \alpha_1 + \alpha_2) - 3 \left(4 m \rho^2 -5 q^2 - 3 m^2 \right) \alpha_3 \right) \, .
\end{align}
Restoring factors of $M_{\text{P}}$, we get
\begin{equation}
    \begin{split}
    &\Delta v^{\,\text{geom}} = \frac{\sqrt{2} \pi \left( 24 M^3 - 6 M_{\text{P}}^3 M Q^2 + 5 M_{\text{P}}^3 \rho^2 (2 M^2 - M_{\text{P}}^3 Q^2) + 24 M_{\text{P}}^6 M \rho^4)\right)}{8 \epsilon M_{\text{P}}^9 \, \rho^5},\\
    &\Delta v^{(1)} = \Delta v^{\,\text{geom}} + \frac{15 \sqrt{2} \pi Q^2}{ \epsilon \rho^5} \alpha_2 + \frac{\sqrt{2} \pi}{\epsilon M_{\text{P}}^6 \, \rho^5}\left( 12 M_{\text{P}}^3 M \rho^2 - \frac{15}{2}M_{\text{P}}^3 Q^2 - 24 M^2 \right) \alpha_3 \, ,\\
    &\Delta v^{(2)} = \Delta v^{\,\text{geom}} + \frac{30 \sqrt{2} \pi Q^2}{\epsilon  \rho^5}(2 \alpha_1 + \alpha_2) - \frac{\sqrt{2}\pi}{ \epsilon M_{\text{P}}^6 \, \rho^5} \left( 24 M_{\text{P}}^3 M \rho^2 - 30 M_{\text{P}}^3 Q^2 - 36 M^2 \right) \alpha_3 \, .
    \end{split}
\end{equation}

Even at leading (two-derivative) order, this expansion leads to complicated polynomials of $ 1 / \rho$. We have kept terms up to $\rho^{-5}$ because that is where $\alpha_1$ and $\alpha_2$ begin to contribute. Let's consider two different limits:
\begin{itemize}
    \item When the black hole is not charged, we have: 
    \begin{align}
        \Delta v^{(1)} = \frac{3 M \sqrt{2} \pi}{ \epsilon \rho M_{\text{P}}^3} \left( 1 + \frac{5 M }{12 M_{\text{P}}^3 \rho^2} + \frac{4 }{ \rho^2} \alpha_3 \right)  \, , \\
        \Delta v^{(2)} = \frac{3 M \sqrt{2} \pi}{ \epsilon  \rho M_{\text{P}}^3} \left( 1 + \frac{5 M }{12 M_{\text{P}}^3 \rho^2} - \frac{8 }{ \rho^2} \alpha_3 \right)  \, .
    \end{align}
We see that the middle term represents a correction to the CEMZ bound 
\cite{Camanho:2014apa}. Without it, the EFT breakdown occurs when $\rho \sim \Lambda^{-1}$, assuming that $\alpha_3 \sim \Lambda^{-2}$. 
 \end{itemize}
\begin{itemize}
    \item In the extremal limit $Q = M / M_{\text{P}}^{3/2}$ we have
    \begin{equation}
    \label{largerhoext}
    \begin{split}
    & \Delta v^{\,\text{geom}} = \frac{\sqrt{2} \pi \left( 18 M^3  + 5 M_{\text{P}}^3 \rho^2  M^2 + 24 M_{\text{P}}^6 M \rho^4)\right)}{8 \epsilon M_{\text{P}}^9 \rho^5} \, ,\\
    &\Delta v^{(1)} = \Delta v^{\,\text{geom}} + \frac{1}{\epsilon} \left[ \frac{15 \sqrt{2} \pi M^2}{ M_{\text{P}}^3 \rho^5} \alpha_2 + \frac{\sqrt{2} \pi}{M_{\text{P}}^6 \rho^5}\left( 12 M_{\text{P}}^3 M \rho^2 - \frac{63}{2}M^2 \right) \alpha_3 \,  \right] , \\
    &\Delta v^{(2)} = \Delta v^{\,\text{geom}} +  \frac{1}{\epsilon} \left[ \frac{30 \sqrt{2} \pi M^2}{M_{\text{P}}^3 \rho^5}(2 \alpha_1 + \alpha_2) - \frac{\sqrt{2}\pi}{M_{\text{P}}^6 \rho^5} \left( 24 M_{\text{P}}^3 M \rho^2 - 66 M^2 \right) \alpha_3 \right] \, .
    \end{split}
\end{equation}
We stress once again that extremality is possible in this large distance regime, while it was not reachable by working in the scaling limit. This differs from the near-horizon extremal limit that we shall discuss below. In the near horizon case, we shall see that the higher-derivative terms and the two-derivative terms scale with the same power of $\rho - \rho_h$. 
\item 
Massless limit: One can also consider the case where $Q >> M$. In this limit, we find 
\begin{equation}
    \begin{split}
    &\Delta v^{(1)} = \frac{\sqrt{2} \pi Q^2}{8 \epsilon  M_{\text{P}}^3 \rho^5} \left( -  5 \rho^2 + 120  M_{\text{P}}^3\alpha_2 - 60     \alpha_3 \right)\\
    &\Delta v^{(2)} = \frac{\sqrt{2} \pi Q^2}{8 \epsilon  M_{\text{P}}^3 \rho^5} \left( -5 \rho^2 + 240   M_{\text{P}}^3 (2 \alpha_1 + \alpha_2) +  240  \alpha_3 \right)
    \end{split}
\end{equation}
We see here that the time-delay would be negative, even at the two-derivative level, indicating that this limit cannot really be taken given our other assumptions.
\end{itemize}

\subsubsection{Near horizon limit}
\label{NHsubsection}

Next, we want to examine the time delays in the opposite regime, where the impact parameter is small, and in particular, it is close to the black hole horizon $\rho_h$. 
We will consider two ways to approach the horizon and comment on the structure of the time delays in each case. 
First, we take the impact parameter to be
    \begin{equation}
\rho = \rho_h(1+\lambda) \quad \text{with}
\quad \lambda <<1 \, ,
\end{equation}
and expand 
perturbatively in
 $\lambda$. We find the following time delays,
\begin{equation} \label{near_hor}
    \begin{split}
    \Delta v^{(1)} &= \frac{1}{\epsilon \lambda^{3/2}} 
    \frac{\pi \sqrt{2}}{(m^{2} - q^{2})^{3/4}}
    \Biggl[ \frac{ 2 m (m + \sqrt{m^{2} - q^{2}})- q^{2}}{16 } 
    + 6  (m - \sqrt{m^{2} - q^{2}})
    (2 \alpha_{1} + \alpha_{2}) \\& + \sqrt{m^{2} - q^{2}} \, \alpha_{3} \Biggr] \, ,\\
    \Delta v^{(2)} &= \frac{1}{\epsilon  \lambda^{3/2}}  \frac{\pi \sqrt{2}}{(m^{2} - q^{2})^{3/4}}\Biggl[  \frac{2 m (m + \sqrt{m^{2} - q^{2}})- q^{2} }{16 } + 18  (m - \sqrt{m^{2} - q^{2}}) (2 \alpha_{1} + \alpha_{2}) \\&+ (2m - 5 \sqrt{m^{2} - q^{2}}) \, \alpha_{3} \Biggr] \, ,
        \end{split}
\end{equation}
where we have only kept the leading $1/\lambda$ contributions. 
For an uncharged black hole, these expressions simplify significantly: 
    \begin{equation}
        \Delta v^{(1)} = \frac{\pi \sqrt{2} \, (m + 4 \alpha_3)}{4 \epsilon \sqrt{m} \,  \lambda^{3/2}}
        [1+ \mathcal{O}(\lambda)] \, , \qquad \Delta v^{(2)} = \frac{\pi \sqrt{2} \, (m-12 \alpha_3)}{4 \epsilon \sqrt{m} \,  \lambda^{3/2}} [1+ \mathcal{O}(\lambda)] .
        %+ \frac{13 \, \sqrt{2}\pi \, (m + 4 \alpha_3)}{16 \sqrt{m}\, \lambda^{1/2}}
    \end{equation}
We find that the time-delay diverges in the near-horizon limit. The most important feature to notice is that the contributions from the geometry and those from the higher-derivative corrections come in \emph{at the same order} in $\lambda$ in this limit, unlike the case of the scaling limit (\ref{neutraldelays}), where the higher-derivative terms were suppressed by the impact parameter. This feature is true independently of the values for the mass and charge of the black hole, and is a consequence of working in 
the near horizon limit. 

At this point, we should note that this expansion is valid and well-behaved as long as
we can think of $\lambda$ as being small.
However, 
for an extremal black hole, we have $\rho_h=0$ and 
$\lambda \equiv \frac{\rho - \rho_{h}}{\rho_{h}}$ 
is no longer perturbative. We will come back to this point further down, but in the meantime, we will assume that this expansion is strictly valid away from extremality.
Finally, restoring units, the time delays given in (\ref{near_hor}) take the form:
\begin{equation}
    \begin{split}
    \Delta v^{(1)} &=  
    \frac{\pi \sqrt{2} \, M_{\text{P}}^{9/2}}{\epsilon \, \lambda^{3/2} \, 
    (4 M^{2} - 2 M_{\text{P}}^{3} Q^{2})^{3/4}} \Biggl[
    \frac{ 4 M \left(2 M + \sqrt{4 M^{2} - 2 M_{\text{P}}^{3} Q^{2}} \, \right)- 2 M_{\text{P}}^{3} Q^{2} }{16 M_{\text{P}}^{6} } \\
    &+ 3   \left(2 M - \sqrt{4 M^{2} - 2 M_{\text{P}}^{3} Q^{2}}\right) (2 \alpha_{1} + \alpha_{2})
    + M_{\text{P}}^{-3} \sqrt{4 M^{2} - 2 M_{\text{P}}^{3} Q^{2}} \alpha_{3} \Biggr],\\
    \Delta v^{(2)} 
    &= 
    \frac{\pi \sqrt{2} M_{\text{P}}^{9/2} }{\epsilon \, \lambda^{3/2} \, 
    (4 M^{2} - 2 M_{\text{P}}^{3} Q^{2})^{3/4}}
    %%%
   \Biggl[  \frac{  4 M \left(2 M + \sqrt{4 M^{2} - 2 M_{\text{P}}^{3} Q^{2}}\right) - 2 M_{\text{P}}^{3} Q^{2}}{16 M_{\text{P}}^{6} } \\&+ 9 \left(2 M - \sqrt{4 M^{2} - 2 M_{\text{P}}^{3} Q^{2}}\right) (2 \alpha_{1} + \alpha_{2}) + M_{\text{P}}^{-3} \left(4 M - 5 \sqrt{4 M^{2} - 2 M_{\text{P}}^{3} Q^{2}}\right) \alpha_{3} \Biggr]\,.
    \end{split}
\end{equation}
%%%%%
Now, if we \emph{naively} go to the extremal limit $m=q$ by looking at the leading terms in the expansion in 
\begin{equation}
\label{zeta}
\zeta = \sqrt{1 - \frac{q^{2}}{m^{2}}} \, ,
\end{equation}
we obtain the following results 
\begin{equation}
\label{badNH}
    \begin{split}
    &\Delta v^{(1)} = \frac{\pi}{8 \epsilon \sqrt{2}\, m^{1/2}} \frac{1}{\lambda^{3/2}}\frac{1}{\zeta^{3/2}} \left( m + 96 (2 \alpha_{1} + \alpha_{2}) \right),\\
    &\Delta v^{(2)} = \frac{\pi}{8 \epsilon \sqrt{2}\, m^{1/2}} \frac{1}{\lambda^{3/2}}\frac{1}{\zeta^{3/2}} \left(  m + 288  (2 \alpha_{1} + \alpha_{2}) + 32 \alpha_{3}  \right).
    \end{split}
\end{equation}
However, as we explained above, this is \emph{not} a well-defined expansion, since at extremality the parameter $\lambda$ is no longer perturbatively small.
Thus, we should identify a better way of examining the time delay in the extremal limit.

\paragraph{Extremality first} A more reliable way to probe the near horizon region of an extremal black hole may be to take \emph{first} the extremal limit $m=q$, and then 
choose the impact parameter to be small (close to $\rho_h\sim0$) in an appropriate manner.
Indeed, we will now go to the extremal limit by looking at the leading term in the expansion (\ref{zeta}) in small $\zeta$, and then take the limit where $\rho^2 << m$.
More precisely, expanding in $\kappa\equiv\rho/\sqrt{m}<<1$,
we get the following contributions to the time delays for our two polarizations 
\begin{equation}
    \begin{split}
     &\Delta v^{(1)} = \frac{\pi}{2 \sqrt{2 m} \epsilon\, \kappa^3 \, }\left( m + 96 (2 \alpha_1 + \alpha_2)\right)  [1+\mathcal{O}(\kappa^2)],  \\
     &\Delta v^{(2)} = \frac{\pi}{2 \sqrt{2 m} \epsilon\, \kappa^3 \, }\left( m + 240 (2 \alpha_1 + \alpha_2) + 24 \alpha_3 \right) [1+\mathcal{O}(\kappa^2)] \, .
    \end{split}
\end{equation}
While they agree qualitatively with (\ref{badNH}), the precise numerical factors are different, showing explicitly that the near horizon limit and extremality don't commute.

It's interesting that the first polarization is not sensitive to $\alpha_3$ to this order in $\lambda$ ($\alpha_3$ does appear in the next order in the expansion).  
It's also interesting that, unlike in the pure field theory case, in the near horizon region we only see dependence on $2 \alpha_1 + \alpha_2$ (as well as on $\alpha_3$) and not on the $\alpha_2$ coefficient on its own. This also shows that working in the near horizon region can yield bounds that differ from the ones we read off in the scaling limit. We also see that we cannot recover the pure field theory results in this limit, which is not surprising because we are working near the black hole horizon.

\paragraph{Allowed regions}

We can now examine the implications of our results for arbitrary values of the charge and mass, as well as for extremal solutions.
Indeed, with some additional assumptions, we can get bounds on the Wilson coefficients of the higher-derivative expansion. 
First, in our analysis below we will assume that we can go arbitrarily close to the horizon, while in reality we should keep $\rho - \rho_h \gg 1 / \Lambda$, to ensure that we are in the range of validity of the EFT description. 
We will also need to make some assumptions about the size of the smallest possible black hole in Planck units. 

We examine the bounds coming from the near-horizon region where, as 
we mentioned in Section \ref{NHsubsection}, the contributions from the geometry and those of the higher-derivative terms appear at the same order 
in $\rho$ (i.e., the $\alpha_i$ terms are not suppressed by additional powers of the impact parameter). This is interesting, because it will provide us with a clear way to see the direct competition between the strength of the bounds on the Wilson coefficients, the mass of the black hole and the range of validity of the EFT.  

\begin{figure}
     \centering
     \begin{subfigure}[t]{0.45\textwidth}
         \centering
         \includegraphics[width=\textwidth]{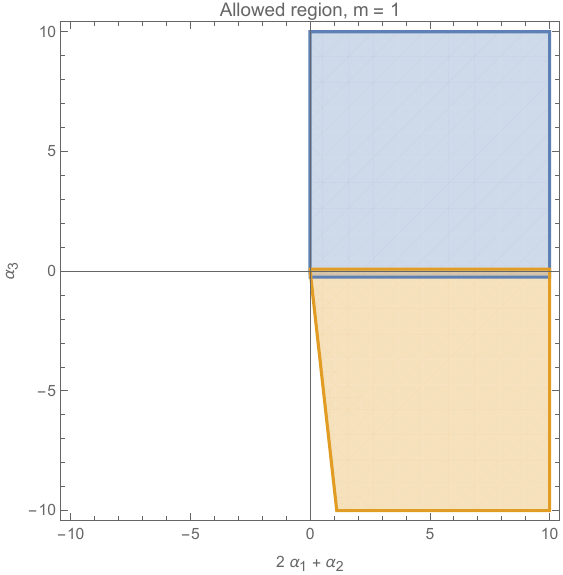}
     \end{subfigure}
     \hfill
     \begin{subfigure}[t]{0.45\textwidth}
         \centering
         \includegraphics[width=\textwidth]{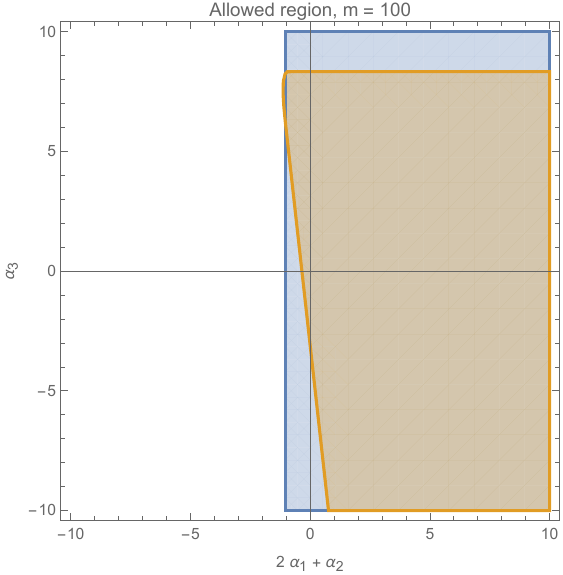}
     \end{subfigure}
     \hfill
        \caption{Allowed regions for the EFT coefficients for (left) $m = 1$ and (right) $m = 100$. The blue / top portion is the allowed region requiring positivity of polarization 1, while the orange / bottom portion is the allowed region from polarization 2.}
        \label{fig:bounds}
\end{figure}

Working with the general expressions for the time delays\footnote{These are too cumbersome to include in the manuscript.} and using $q = m \sqrt{1 - \delta^2}$, where $\delta = 0$ corresponds to extremality, we find that positivity in the near-horizon limit requires
\begin{align}
\begin{split}
\label{eq:nh_ineq}
    0 \ & \leq \ m (1 + \delta)^2 + 96 (2 \alpha_1 + \alpha_2) (1 -\delta) + 16 \alpha_3 \delta \, ,  \\
    0 \ &\leq \ m (1 + \delta)^2 + 288 (2 \alpha_1 + \alpha_2) (1 -\delta) + 16 \alpha_3 (2 - 5\delta) \, .
\end{split}
\end{align}
Then requiring that~\eqref{eq:nh_ineq} holds for all $\delta$ between $0$ and $1$ allows us to plot the allowed region of $2 \alpha_1 + \alpha_2$ vs. $\alpha_3$, as we have done for $m = 1$ and $m = 100$ 
in figure~\ref{fig:bounds}. From these figures we can see that the effect of increasing the minimum allowed $m$ is to zoom in, weakening the bounds. This is simply the statement that the purely gravitational, two-derivative contribution is becoming larger, and thus there is more room for a negative contribution coming from the higher-derivative corrections.
Thus, in the limit of large $m$, everything is allowed, while in the limit of small $m$, we find that only $\alpha_3 = 0$, $2 \alpha_1 + \alpha_2 \geq 0$ is allowed, effectively recovering the non-gravitational bounds. Naively, this would 
tell us that the strongest bounds will come 
from the lighest black hole. 
However, considering small $m$ is not really physical, as we explain directly below.

\paragraph{EFT validity}

We are considering a number of different expansions, so let us try to be explicit about them here. First, we have the large boost expansion, $\gamma \to \infty$. This is the first limit we take, meaning that $\gamma$ is the largest number in the problem. We don't expect this to cause any problems.

The next consideration is the EFT expansion. In particular, we have considered the four-derivative corrections to Einstein-Maxwell theory, but there are terms with more fields (\textit{e.g.} $F^6$) and/or more derivatives (\textit{e.g.} $(\nabla^2 F^2)(F^2)$) which will exist as well. 

One question is how this relates to the near-horizon expansion, where we have taken $\rho$ very close to $\rho_h$. We have seen that the two- and four-derivative terms scale with the same power of $\rho - \rho_h$, but have different powers of $m$. In fact, we have explicitly verified that the coefficient $\tilde \alpha$ of the $(F^2)^3$ corrections also scales with the same power of $\rho - \rho_h$ and is suppressed by an additional power of $m$. 
 
Thus, it is natural to expect that near the horizon every
additional two-derivative contribution is suppressed by an additional power of $m$. We leave verifying this to future work, but if true, it gives a clear indication of the validity of the EFT: we cannot make $m$ too small relative to the EFT scale. In particular, we can write the time delay in the near-horizon limit at extremality (ignoring numbers) as
\begin{align}
    \Delta v = m + (2 \alpha_1 + \alpha_2) + \tilde \alpha / m.
\end{align}
Let us restore the appropriate powers of $M_{\text{P}}$ and define $\alpha_i = a_i / \Lambda^5$ and $\tilde \alpha = \tilde a / \Lambda^{10}$ for some order-one numbers $a_1$, $a_2$, and $\tilde a$. Then we find
\begin{align}
    \Delta v = \frac{M}{ M_{\text{P}}^2} \left(1 + (2 a_1 + a_2) \left(\frac{M_{\text{P}}^6}{ M \Lambda^5}\right) + \tilde a \left(\frac{M_{\text{P}}^6}{ M \Lambda^5}\right)^2 + ... \right) .
\end{align}
Our conjecture is that, in the near-horizon limit, the time delay is a series in $M_{\text{P}}^6 / M \Lambda^5$, therefore the EFT is only valid if we consider the black hole size to satisfy
\begin{align}
    \frac{M}{M_{\text{P}}} \gg \left( \frac{M_{\text{P}}}{\Lambda}\right)^5 \, .
\end{align}
%%%%%%%
Another way to state this is that to preserve causality, we want
\begin{align}
    2 a_1 + a_2 \gtrsim - \left( \frac{M \Lambda^5}{M_\text{P}^6} \right) \, .
\end{align}
We see that the smaller we take $M$, the stronger a bound we find as a result. Keep in mind, however, that the number appearing on the RHS has to be large for the validity of the perturbative expansion, so the number bounding the negativity of the dimensionless parameters $a_1$ and $a_2$ cannot be small. Our bound, rather, is a parametric statement about the size of $a_1$ and $a_2$.

Let us compare these results with what one would get from inspecting 
the large $\rho$ regime. We will focus on the extremal 
case, where the time delays are given by (\ref{largerhoext}).
The time delays for $a_i$ (with $i=1,2$) can be rewritten in the schematic form, neglecting the $\alpha_3$ dependence for simplicity,

\begin{align}
         a_i \gtrsim - \frac{ M \Lambda^5}{M_\text{P}^6}  \left( \frac{M_{\text{P}}^6 }{M^2} \rho^4 +  \frac{M_{\text{P}}^3 }{M} \rho^2 + 1  \right) \,.
\end{align}
In this limit, we cannot read off any (even a parametric) bound, because the expansion was obtained by assuming large $\rho$, and the leading term appearing in the lower bound on $a_2$ is order $\rho^4$. This is the source of our claim that the strongest bounds are obtained by considering the near-horizon limit.

\section{Conclusion}
\label{sec:Conclusion}

In this paper, we have computed bounds on Maxwell and Einstein-Maxwell theory in five dimensions using causality on shockwave backgrounds. In the case of Maxwell theory, \textit{i.e.} pure QFT, this gives positivity bounds on the four-derivative corrections, while when gravity is included, we find the familiar pattern that the field theory bounds are weakened by the contribution of gravity -- the pure QFT bounds $2 \alpha_1 + \alpha_2 > 0$ and $\alpha_2 > 0$ are recovered in the $M_{\text{P}} \to \infty$ limit. The full gravitational theory, where the boosted sources are charged black holes, gives a number of limits to play with, including the scaling limit in which $m_0 = \gamma m$ and $q_0^2 = \gamma q^2$ are kept finite as $\gamma \to \infty$, as well as the near-horizon and the large impact parameter 
limits. In the latter two cases we identify where the EFT breaks down -- beyond that point the time delay is sensitive to the full UV completion.

A non-trivial technical point of our work is the elimination of the scaling limit. In the original derivation of shockwaves from boosted black holes \cite{Aichelburg}, the mass was scaled with the charge to recover a finite time delay in the infinite boost limit. Adding a charge to the black hole makes this procedure harder to interpret: while the scaled mass becomes the momentum in the infinite boost limit, the scaled charge does not correspond to any recognizable quantity, and does not make sense in the case that the charge is quantized. Also, the scaling limit does not allow one to examine extremal black holes, since the mass and the charge are scaled differently. 
In this paper, we have handled this issue by allowing the time delay to diverge and imposing positivity on the leading divergence. Thus, for a very large boost the time delay will be a large positive, rather than a large negative, number.
Our analysis also suggests that the most stringent causality bounds on the coefficients $\alpha_1$ and $\alpha_2$ come from probing the near horizon  region of the smallest black holes. This is because the four-derivative terms appear at the same order as the two-derivative terms. However, as one tries to make the bounds more stringent, the EFT quickly breaks down, as expected.
Still, this may provide guidance for connecting with swampland studies.

Finally, let us comment on some future directions. One idea would be to extend these calculations to Anti de Sitter and de Sitter spacetimes. Although this will introduce some technical complications (it may be tricky to isolate the correct polarizations), the calculation should proceed in the same way. 
Computing the time delays in Anti de Sitter will allow us to interpret our bounds in terms of the holographic dual, with possible implications for transport and specifically (for the case of $F^4$ operators) the conductivity of strongly correlated electron systems. 
Causality on bulk shockwaves can be used to derive bounds on the OPE in the Regge limit \cite{Afkhami-Jeddi:2017rmx} -- it would be interesting to explore some of these ideas using the concrete solutions that come from boosted Reissner-Nordstr\"om black holes. In de Sitter space it is known that the time delay from shockwaves in Einstein gravity is negative -- in this case, it is not at all obvious what sort of shift the higher-derivative corrections should imply. It will be interesting to see if corrections which are positive in flat space, such as the $F^4$ terms with positive $\alpha_2$ and $2 \alpha_1 + \alpha_2$, will lead to positive or negative corrections in de Sitter space.

As we mentioned in the introduction, one motivation for studying corrections to Einstein-Maxwell theory is to establish the black hole WGC \cite{Kats:2006xp}, which says that the four-derivative correction to the mass of an extremal black hole should be positive, or to related conjectures on the entropy \cite{Cheung:2018cwt, Goon:2019faz, Cremonini:2019wdk, McPeak:2021tvu} or force between identical black holes  \cite{Heidenreich:2020upe, Cremonini:2021upd, Etheredge:2022rfl}. The issue is that in $d> 4$, these quantities get a contribution from $R_{\mu \nu \rho \sigma} R^{\mu \nu \rho \sigma}$, which can be canceled in 4d by subtracting the topological Gauss-Bonnet term from the action. Therefore in $d > 4$, there are 4, rather than 3 four-derivative corrections to consider, and photon scattering on charged backgrounds can only access terms which have powers of $F^2$. Perhaps by considering both photons and gravitons, one would find the combination of coefficients relevant for the WGC, which is given (in a slightly different basis) in equation (S20) of \cite{Hamada:2018dde}. However, the same fundamental issue -- that gravity weakens the bounds and prevents us from proving the conjecture -- would likely be present in that case. See \cite{Henriksson:2022oeu} for a discussion.

A general open question is how far causality bounds can go towards implying positivity of higher-derivative operators, and how they might compare to the bounds imposed by dispersion relations. Causality is believed to imply maximal analyticity, which is an assumption on the amplitudes used in deriving the dispersive bounds. However, the causality bounds of the type derived in this paper are intrinsically classical, so it remains plausible that they will not be as strong as the bounds coming from the requirements of a unitary causal S-matrix. Recent work  \cite{CarrilloGonzalez:2022fwg, CarrilloGonzalez:2023cbf} has systematically compared the bounds coming from causality and dispersion relations, and the set of backgrounds considered in that paper were not enough to derive all the constraints coming from dispersion relations. In general, the question of whether the bounds are equivalent remains open.

Along similar lines, it would be interesting to see how to establish bounds on subleading higher-derivative operators, like six-derivative corrections to Maxwell theory or scalar field theory. Part of the issue with this computation is that there is no systematic way of decoupling different derivative orders, so it is not clear in our language how to derive bounds which appear obvious in the dispersion relation language, such as the positivity of certain 8-derivative operators (\textit{e.g.} $g_4$ in \cite{Caron-Huot:2020cmc}).

A more ambitious goal would be to adapt our methods to study bounds on six- or higher-point coefficients appearing in the action. This is a pressing problem that has not been seriously addressed in the literature due to numerous technical problems. Here classical solutions may be of use because they are similarly complex for four and higher-point interactions, unlike the amplitudes, which have a huge explosion of kinematics as point order is increased. Still, this suffers a similar problem as subleading derivative operators: it is not obvious how to find any bound that is sensitive to six-point operators without having to consider eight-point, ten-point, and so on. 

Another interesting question we plan to explore systematically is what the focusing theorem \cite{Hartman:2022njz} says about higher-derivative operators. The focusing theorem is a consequence of the null energy condition and roughly says that parallel light rays converge in theories of gravity. In \cite{Hartman:2022njz} this statement was turned into a concrete condition on the Laplacian of the time delay. Thus, it would be very interesting to apply this condition to the time delays derived in this paper and in \cite{Cremonini:2023epg}, to try to derive stronger bounds on the coefficients of Einstein-Maxwell theory.

A final interesting development has been the relation of shockwave backgrounds to the gravitational memory effect. A version of this story was already understood by Dray and 't Hooft \cite{Dray:1984ha}, but recently the relationship has been given a more concrete form, as the shockwave metrics like those in this paper were shown to be related to the Bondi metrics describing gravitational memory \cite{He:2023qha, He:2024vlp}. Causality requires that the time delay is positive, and there may be even more requirements along the lines of the focusing conjecture discussed above. It would be interesting to translate this into a constraint on the memory effect, potentially leading to a drastic reinterpretation of the causality bounds in this paper in the context of soft theorems and asymptotic symmetries. We leave these questions for future work. 

\section*{Acknowledgments}

We would like to thank Simon Caron-Huot, Miguel Correia, Gary Shiu and 
Jan Pieter van der Schaar for useful conversations and correspondences. 
This work has received funding from the European Research Council (ERC) under the European Union's Horizon 2020 research and innovation program (grant agreement no.~758903). 
The work of S.C.\,,\,M.M. and M.R. is supported in part by the NSF grant
PHY-2210271. 
M.R. and M.M. acknowledge the support of the Dr. Hyo Sang Lee Graduate Fellowship from the College of Arts and Sciences at Lehigh University. 

\appendix
\addtocontents{toc}{\protect\setcounter{tocdepth}{1}}

\section{Useful identities}

We write here some of the identities that we used to compute the time delay experienced by the probe particle in the infinite boost limit,

\begin{equation} \label{coeff}
\begin{split}
\lim _{\epsilon \rightarrow 0} \frac{\epsilon^{n-\frac{1}{2}}}{\left(u^2+2 \epsilon \rho^2\right)^n}&=\frac{\Gamma\left(n-\frac{1}{2}\right)}{\Gamma(n)} \frac{\pi^{1 / 2}}{(\sqrt{2} \rho)^{2 n-1}} \delta(u)  \\[2mm]
   \lim _{\epsilon \rightarrow 0} \frac{\epsilon^{n-\frac{1}{2}}}{\left(u^2+ \epsilon (m+q+2\rho^2)\right)^n}&=\frac{\Gamma\left(n-\frac{1}{2}\right)}{\Gamma(n)} \frac{\pi^{1 / 2}}{\Big( \sqrt{2 \rho^2+m+q}\Big)^{2 n-1}} \delta(u) \\[2mm]
    \lim _{\epsilon \rightarrow 0} \frac{\epsilon^{n-\frac{1}{2}}}{\left(u^2+ \epsilon (m-q+2\rho^2)\right)^n}&=\frac{\Gamma\left(n-\frac{1}{2}\right)}{\Gamma(n)} \frac{\pi^{1 / 2}}{\Big( \sqrt{2 \rho^2+m-q}\Big)^{2 n-1}} \delta(u)  \\[2mm]
    \lim _{\epsilon \rightarrow 0} \frac{\epsilon^{n-\frac{1}{2}}}{\left(u^2+ \epsilon (\sqrt{m^2-q^2}+2\rho^2)\right)^n}&=\frac{\Gamma\left(n-\frac{1}{2}\right)}{\Gamma(n)} \frac{\pi^{1 / 2}}{\Big( \sqrt{2 \rho^2+\sqrt{m^2-q^2}}\Big)^{2 n-1}} \delta(u) \\[2mm]
    \lim _{\epsilon \rightarrow 0} \frac{\epsilon^{n-\frac{1}{2}}}{\left(u^2+ \epsilon (-\sqrt{m^2-q^2}+2\rho^2)\right)^n}&=\frac{\Gamma\left(n-\frac{1}{2}\right)}{\Gamma(n)} \frac{\pi^{1 / 2}}{\Big( \sqrt{2 \rho^2-\sqrt{m^2-q^2}}\Big)^{2 n-1}} \delta(u)\,.
   \end{split}
\end{equation}

\section*{}
\bibliography{cite.bib}
\bibliographystyle{JHEP.bst}
\end{document}